\documentclass[twocolumn,showpacs,prb]{revtex4}
\usepackage{graphicx}%
\usepackage{dcolumn}
\usepackage{amsmath}
\usepackage{amssymb}
\usepackage{bm}
\makeatletter
\def\btt#1{\texttt{\@backslashchar#1}}%
\DeclareRobustCommand\bblash{\btt{\@backslashchar}}%
\makeatother

\begin{document}

\preprint{BC8222.tex}

\title{Crystal and magnetic structure of the La$_{1-x}$Ca$_{x}$MnO$_{3}$ compound $(x=0.8,0.85)$.}
\author{M. Pissas, G. Kallias}
\affiliation{Institute of Materials Science, NCSR,  Demokritos, 153 10 Aghia Paraskevi, Athens, Greece}
\author{M. Hofmann, D. M. T\"{o}bbens}
\affiliation{Berlin Neutron Scattering Center, Hahn-Meitner-Institut Glienicker Str. 100 D-14109, Berlin, Germany}
\date{\today }

\begin{abstract}
We studied the crystal and magnetic structure of the La$_{1-x}$Ca$_{x}$MnO$_{3}$
compound for $x=0.8$ and $x=0.85$. At $T=300$ K both samples are paramagnetic
with crystallographic symmetry $Pnma$. At low temperatures they undergo a
monoclinic distortion from orthorhombic $Pnma$-type structure with
$a_p\sqrt{2}\times 2a_p\times a_p\sqrt{2}$ to a monoclinic structure with
($a_p\sqrt{2}\times 2a_p\times a_p\sqrt{2}$, $\beta=90+\varepsilon\sim 91.4^{\rm o}$)
and $P2_1/m$ space group below $T_N$.
The onset of the structural transformation coincides with the development of
the $C$-type long range antiferromagnetic order with propagation vector
${\bf k}=(\frac{1}{2},0,\frac{1}{2})$.
The monoclinic unit cell allowed us to determine the direction of the Mn magnetic moment with
respect to the crystallographic axes: it is perpendicular to
the propagation vector, ${\bf m}\perp {\bf k}=(\frac{1}{2},0,\frac{1}{2})$. The amplitude
of the ordered magnetic moment at $T=1.6$ K is found to be $2.53(2)$ and $2.47(2)\mu _{B}$ for
$x=0.8$ and $0.85$, respectively.
\end{abstract}
% 75.50.Ee Antiferromagnetics
%64.70.Kb Solid-solid transitions (see also 61.50.K
%     Crystallographic aspects of phase
%     transformations, pressure effects; 75.30.K
%     and 77.80.B for magnetic and ferroelectric
%     transitions, respectively; for material
%     science aspects, see 81.30)
\maketitle
\section{Introduction}

The interest in the mixed perovskites\cite{wollan55,goudenough55} La$_{1-x}$%
\-Ca$_{x}$\-MnO$_{3}$ has been renewed \cite
{schiffer95,kuwahara95,tomioka95,lynn96,ramirez96,satpathy96,millis98,nagaev96,majumdar98,moreo99,varelogiannis,ramirez97,rao,pissas97}
in connection with the discovery of colossal magnetoresistance for $x\sim
0.33$. \cite{helmolt93} For the high doping range with $x>0.5$, the Mn$^{3+}$
$e_{g}$ electrons become localized at low temperatures with a concomitant
anisotropic bond reorganization in the oxygen environment of the manganeses.
This cooperative atomic rearrangement can change the symmetry or not,
depending upon the particular electronic concentration, orbital occupancy,
average size and distribution in the La site. Thus, when the temperature is
lowered through a transition temperature $T_{co}$, charge and orbital
ordering occur. Up to now two distinct cases of charge-ordering have been
reported for $x=0.50$ and $x=0.67$ both by transmission electron microscopy
and synchrotron x-ray and neutron powder diffraction. \cite
{radaelli97,chen96,mori98,mori98a,radaelli99,fernandez99}

At half-doping ($x=0.50$), the system undergoes a phase transition from
paramagnetic (insulating) to ferromagnetic-metallic (FM) phase at onset
temperature $T_{c}=234$ K and then upon cooling to an
antiferromagnetic-insulating (AFM) phase at $T_{N}^{{\rm c}}=163$ K. \cite
{roy98} The antiferromagnetic phase presents charge and spin ordered
structure, called CE type\cite{wollan55} where real space ordering of Mn$%
^{3+}$ and Mn$^{4+}$ takes place. The basic characteristic of $x=1/2$ phase,
at low temperature, is the one after the other ordering of the Mn$^{3+}$ and
Mn$^{4+}$ ions along the $a\approx \sqrt{2}a_{p}$ ($Pnma$ setting, $a_{p}$
is the pseudo cubic unit cell parameter of the ideal cubic perovskite
structure) leading to a superstructure with a propagation vector ${\bf k}%
=(2\pi /a)(1/2,0,0)$. \cite{radaelli97,chen96,mori98,mori98a,kallias00,huang00}

For $x=0.67$ pioneer work of Radaelli\cite{radaelli99,fernandez99}
and collaborators has shown a crystallographic charge-ordered\cite
{chen96,mori98,mori98a} and magnetic superstructure. The
antiferromagnetic structure is noncollinear with the $a$ lattice
parameter to be tripled and the $c$ lattice parameter to be
doubled with respect to the average crystallographic unit cell
$Pnma$ setting. The crystallographic structure below the
charge-ordering temperature ($T_{{\rm CO}}\approx 260$ K) is
characterized by ordering of the $d_{z^{2}}$ orbitals of the
Jahn-Teller-distorted Mn$^{3+}$O$_{6}$ octahedron in the
orthorhombic $ac$ plane, and the appearance of superlattice peaks
in the x-ray patterns corresponding to a tripling of the $a$ axis
lattice parameter. The refinement has revealed ordering of the
Mn$^{3+}$ cations in sites as far apart as possible in the $ac$
plane ''Wigner-crystal'' model and transverse displacements of the
Mn$^{4+}$O$_{6}$ octahedra in the $c$ direction. A recent study on
$x=2/3$ \cite{wang00} combining transmission electron microscopy
with high resolution synchrotron powder x-ray diffraction data
also supports the ''Wigner crystal'' model.

Moreover, transmission electron microscopy studies in La$_{1-x}$Ca$_{x}$MnO$%
_{3}$ with $x=0.50,0.625,0.67,0.75$ and $0.80$ have shown superlattice
reflections and their analysis pointed to a linear dependence between the
magnitude of the modulation wave vector ${\bf q}_{s}$ along the $a$ axis and
$x$, that is, $q_{s}=1-x$. \cite{li99} However, the superlattice reflections
of the charge-ordering for $x=0.8$ \cite{chen96,li99} were noticeably
broader than lower compositions, indicating a very short coherence length.

Given the interplay of structural, transport, and magnetic properties on
these highly correlated electron systems it is extremely important to
provide information on the crystal and magnetic structure for $x$ values
above $0.75$, for which the existing literature is rather limited.\cite
{chen96,li99}

In the present paper we discuss the crystal and magnetic structure for
doping levels near the end members of the La$_{1-x}$Ca$_{x}$MnO$_{3}$
homologous series using elastic neutron diffraction data from
polycrystalline La$_{1-x}$Ca$_{x}$MnO$_{3}$ samples with $x=0.8$ and $0.85$.
An important question that is been posed is whether charge ordering is
actually present or not.

%-------------------------------------------------------------------
\begin{figure}[tbp] \centering%
\includegraphics[angle=0,width=7.5cm]{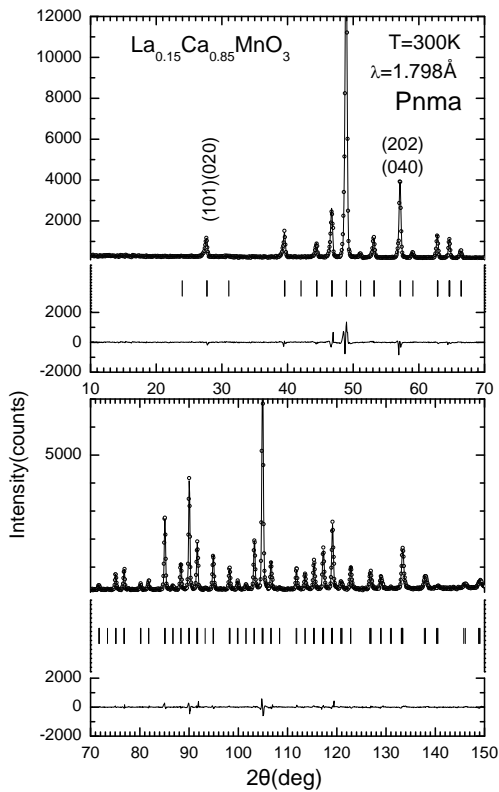}
\caption{
Rietveld refinement for La$_{0.15}$Ca$_{0.85}$MnO$_{3}$ at 300 K
($\lambda=1.798$\AA).The observed data points are indicated with open circles,
while the calculated pattern is shown as a continuous line. The
positions of the reflections are indicated with vertical lines below the pattern.}
\label{fig1}%
\end{figure}%
%---------------------------------------------------------

\section{Experimental details}

La$_{1-x}$Ca$_{x}$MnO$_{3}$ samples with $x=0.80$ and $x=0.85$ were prepared
by thoroughly mixing high purity stoichiometric amounts of CaCO$_{3}$, La$%
_{2}$O$_{3}$, and MnO$_{2}$. The mixed powders reacted in air in
temperatures up to 1400$^{\text{o}}$C for several days with intermediate
grindings. Finally the sample was slowly cooled to room temperature. Neutron
diffraction data were collected on the E6 and E9 diffractometers of the
research reactor BER II in Berlin. The neutron powder diffraction
experiments as a function of temperature in the low angle range were
performed in diffractometer E6 using a wavelength $\lambda =2.44$\AA ((002)
reflection of a pyrolytic graphite monochromator). For crystal structure
refinement, data were collected on the E9 diffractometer with wave length $%
\lambda =1.798$\AA\ ((511) reflection of a vertically focusing Ge
monochromator), with collimation $\alpha _{1}=10^{/}$(in pile collimator), $%
\alpha _{2}=20^{/}$(second collimator after monochromator) and $64\times
10^{/}$ collimators in front of 64 $^{3}$He single detector tubes. \ The
powdered samples were placed in a cylindrical vanadium can ($D=8$ mm)
mounted in an ILL orange cryostat. DC magnetization measurements were
performed in a superconducting quantum interference device (SQUID)
magnetometer (Quantum Design).

\section{Crystal structure}

We performed Rietveld refinements of the nuclear and magnetic structures of
neutron data using the {\sc FULLPROF} program.\cite{fullprof} The
systematically absent reflections of the $x=0.8$, $0.85)$ diffraction
patterns at 300 K are consistent with space group $Pnma$. The crystal
structure at 300 K was refined using as starting model the orthorhombic GdFeO%
$_{3}$ type structural model. First, the scale factor, background, unit cell
parameters and zero-shift errors were optimized. The peak shapes are well
described by function proposed by Finger, Cox and Jephcoat\cite{finger94} in
order to model appropriately the high vertical apertures used in the beam
geometry.\cite{tobbens01} Then, we refined the atomic positions and the
isotropic thermal parameters. In the final step, we refined the oxygen
thermal parameters anisotropically reducing further the refinement agreement
parameters. Figure \ref{fig1} shows as an example the 
%Rietveld plot 
observed and calculated neutron powder diffraction intensity patterns
for the La$_{0.15}$Ca$_{0.85}$MnO$_{3}$ sample at 300 K. The corresponding
structural parameters are reported in Table \ref{table1} and the selected
bond distances in Table \ref{table2}. The structure at this temperature is
orthorhombic with $a>c>b/\sqrt{2}$. Assuming that the Mn site is completely
occupied and refining the occupation factors for La/Ca and O we found the
nominal ones within the standard deviation errors. Judging from the
estimated Mn-O distances, the octahedral distortion is small and can be
considered as slightly tetragonally elongated with four short bonds $\sim
1.91$\AA\ (two in the basal plane ($a-c$-plane) and two apical bond (along $%
b-$axis)) and two longer bonds $\sim 1.92$\AA\ in the basal plane.
%----------------------------------------------------------------------------
\begin{figure*}[tbp] \centering%
\includegraphics[angle=0,width=15cm]{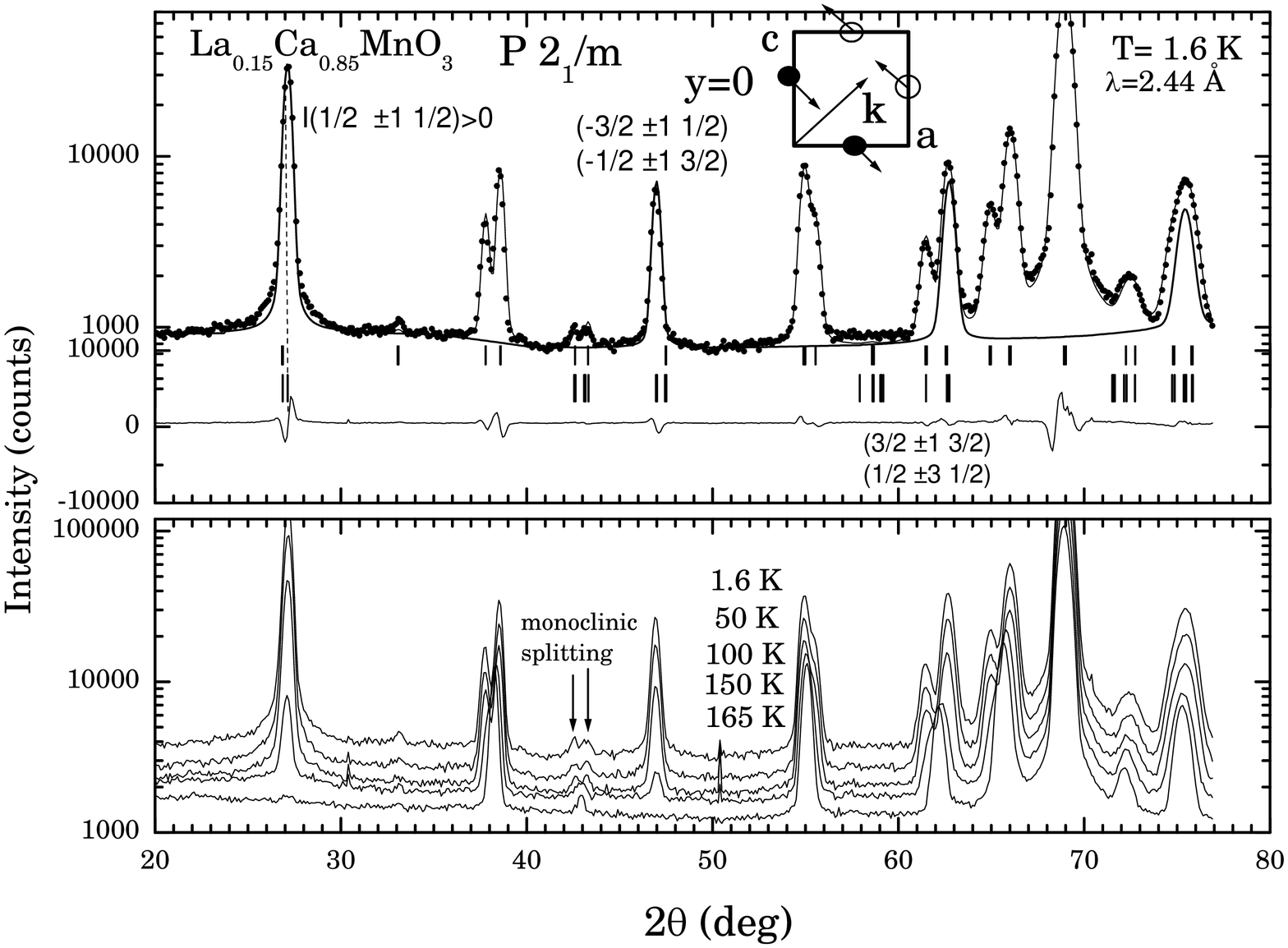}%
\caption{
Observed and calculated neutron powder diffraction intensity patterns
%Rietveld plot 
at $T=1.6$ K and neutron diffraction patterns at $50$ K, $100$ K, $150$ K
and $165$ K of the La$_{0.15}$Ca$_{0.85}$MnO$_{3}$ sample ($\lambda=2.44$\AA).
The observed data points are indicated with solid circles,
while the total structural and magnetic calculated pattern are shown as continuous lines. The
positions of the reflections for the crystal and magnetic structure
employed in the refinements are indicated with vertical lines below the pattern.
The indices refer to the magnetic reflections. The splitting of the (111) reflection
(at 43 degrees) due to the monoclinic distortion can also be seen.
}
\label{fig2}%
\end{figure*}%
%---------------------------------------------------------------------------

The neutron diffraction patterns, below a temperature denoted by $T_{c}$,
indicate a change of symmetry since several extra peaks appear with respect
to the high temperature $Pnma$ space group. Figure \ref{fig2} shows the
neutron diffraction patterns collected with a wavelength $2.44$\AA\ (E6
diffractometer) for the $x=0.85$ sample from $T=1.6$ K up to $165$ K. Close
inspection of the diffraction pattern at $T=1.6$ K shows a splitting of the $%
(111)$ reflection, characteristic for the existence of monoclinic distortion
of the structure with the monoclinic angle $\beta >90^{o}$. In the case of a
monoclinic distortion of the unit cell, one expects a splitting of the
reflections $(111)$ and $(\underline{1}11),$($d^{-2}(\underline{1}%
11)-d^{-2}(111)$ $\propto $ $-\cos \beta $), since the cross term $hl$ in
the expression for the d-spacing for these reflections is non zero.
%-------------------------------------------------
\begin{table}[tbp] \centering%
\caption{La$_{1-x}$Ca$_{x}$MnO$_{3}$ ($x=0.8$ and $0.85$) structural parameters at
$T=300$ and $1.6$ K
as determined from Rietveld refinements based on neutron powder diffraction data of the E9
diffractometer ($\lambda=1.798$\AA ) The space group $Pnma$ (No62) was used for the data at
$T=300$ K. La, Ca and apical oxygen (Oa) occupy the $4c$, $(x,1/4,z)$ site, Mn the 4b $(0,0,1/2)$
site and the plane oxygen (Op) the general $8d$ site. The monoclinic space group $P 2_1/m$
($b$ axis unique No 11) was used for the data at $T=1.6$ K.
The atomic sites for $P 2_1/m$ are: La/Ca1,2, Oa1,2 $2e [x,\frac{1}{4},z]$,
Mn1 $2b [\frac{1}{2},0,0]$, Mn2 $2c [0,0,\frac{1}{2}]$, and Op1,2 $4f [x,y,z]$.
The parameters $U_{ij}$ represent the eigenvalues of the cartesian $U-$ matrix.
Numbers in parentheses are statistical errors referring to the last significant digit.}
\begin{ruledtabular}
\begin{tabular}{ccccc}
$x$ & \multicolumn{2}{c}{0.8} & \multicolumn{2}{c}{0.85} \\
$T$ (K) & 300 & 1.6 & 300 & 1.6 \\
\tableline$a$ (\AA ) & 5.3386(1) & 5.3529(2) & 5.3243(1) & 5.3356(1) \\
$b$ (\AA ) & 7.5376(1) & 7.4648(3) & 7.5193(1) & 7.4596(1) \\
$c$ (\AA ) & 5.3367(1) & 5.3506(2) & 5.3189(1) & 5.3258(1) \\
$\beta ^{{\rm o}}$ & 90 & 91.502(2) & 90 & 91.141(1) \\
La1 $x$ & 0.0240(7) & 0.022(2) & 0.0247(7) & 0.029(1) \\
$z$ & -0.006(1) & -0.005(1) & -0.009(1) & -0.005(1) \\
$B$ (\AA $^{2}$) & 1.13(6) & 0.79(7) & 0.89(6) & 0.70(5) \\
La2 $x$ &  & 0.522(2) &  & 0.528(1) \\
$z$ &  & 0.509(2) &  & 0.509(1) \\
$B$ &  & 0.81(7) &  & 0.70(5) \\
Mn1 $B$ & 0.80(7) & 0.43(7) & 0.49(6) & 0.37(6) \\
Mn2 $B$ & -       & 0.43(7) & - & 0.37(6) \\
O1a $x$ & 0.4921(8) & 0.497(2) & 0.4913(7) & 0.495(1) \\
$z$ & 0.063(1) & 0.061(1) & 0.061(1) & 0.061(1) \\
$U_{11}$ (\AA $^{2}$) & 0.014(4) & 0.006(3) & 0.013(6) & 0.015(1) \\
$U_{22}$ & 0.0041(3) & 0.001(1) & 0.008(7) & 0.002(1) \\
$U_{33}$ & 0.0095(4) & 0.011(1) & 0.010(4) & 0.010(2) \\
O2a $x$ &  & 0.988(2) &  & 0.985(1) \\
$z$ &  & 0.438(1) &  & 0.434(1) \\
$U_{11}$ &  & 0.006(3) &  & 0.015(3) \\
$U_{22}$ &  & 0.001(1) &  & 0.002(1) \\
$U_{33}$ &  & 0.011(1) &  & 0.010(3) \\
O1p $x$ & 0.2827(7) & 0.282(1) & 0.2828(6) & 0.2829(8) \\
$y$ & 0.0414(4) & 0.038(1) & 0.0338(5) & 0.0342(8) \\
$z$ & 0.7155(8) & 0.721(1) & 0.7151(7) & 0.7198(9) \\
$U_{11}$ & 0.0075(3) & 0.022(2) & 0.0001 & 0.011(2) \\
$U_{22}$ & 0.0114(8) & 0.006(1) & 0.01638 & 0.003(1) \\
$U_{33}$ & 0.0202(5) & 0.003(1) & 0.0174 & 0.007(5) \\
O2p $x$ &  & 0.781(1) &  & 0.7842(8) \\
$y$ &  & 0.028(1) &  & 0.0308(6) \\
$z$ &  & 0.779(1) &  & 0.782(1) \\
$U_{11}$ &  & 0.022(2) &  & 0.011(2) \\
$U_{22}$ &  & 0.006(1) &  & 0.003(1) \\
$U_{33}$ &  & 0.003(1) &  & 0.007(1) \\
\tableline$R_{p}$ & 6.86 & 7.7 & 6.9 & 6.60 \\
$R_{wp}$ & 9.25 & 9.9 & 9.7 & 9.21 \\
$R_{B}$ & 6.35 & 4.86 & 5.09 & 4.82 \\
$\mu _{x}$ &  & 1.79(1) &  & 1.73(1) \\
$\mu _{z}$ &  & -1.79(1) &  & -1.73(1) \\
$\mu $ &  & 2.53(2) &  & 2.47(2) \\
$R_{B}(m)$ &  & 6.8 &  & 6.59
\end{tabular}
\end{ruledtabular}
\label{table1}%
\end{table}%
%------------------------------------------------

The structure at $T=300$ K is associated with the GdFeO$_{3}$/CaTiO$_{3}$
type structures (like the samples with different Ca content) and is derived
from the cubic perovskite by a combination of MnO$_{6}$ octahedra tilts
(Glazer \cite{glazer72} notation $Pnma,\,a^{-}b^{+}a^{-}$ ). Looking in
Glazer's classification for an octahedral tilt system which maintains the
tilt system $a^{-}b^{+}a^{-}$ and is compatible with a monoclinic unit cell,
we selected the tilt system $a^{-}b^{+}c^{-}$ to refine the low temperature
monoclinic phase. This tilt system is described with the monoclinic space
group $P2_{1}/m$, which is a subgroup of $Pnma$. In $P2_{1}/m$ space group
there are two non-equivalent sites for La and Mn.\cite{woodward97} Each Mn
site has its own apical oxygen (bond direction mainly along $b$-axis).
Concerning the two plane oxygens at the general position they are shared by
the two Mn sites. 
%change
In order to reduce the free parameters we used the constrains 
$B({\rm La(1)})=B({\rm La(2)})$ and $B({\rm Mn(1)})=B({\rm Mn(2)})$ for La and Mn
temperature factors respectively.
The refinement of neutron diffraction patterns after
taken the above constraints into consideration gives very good Rietveld agreement factors. 
% end change
The same type of monoclinic distortion has been observed in Sm$_{0.15}$Ca$_{0.85}
$MnO$_{3}$ and Bi$_{0.15}$Ca$_{0.85}$MnO$_{3}$ compounds.\cite
{martin99,llobet00} \ Due to the low x-ray scattering factor of oxygen the
experiments done by Zheng et al.\cite{zheng} were not sensitive enough to
find the correct structural transition in La$_{0.17}$Ca$_{0.83}$MnO$_{3}$.

We have also tested the space group $P2_{1}/n$ (No 14) which follows the
tilt system $a^{-}b^{-}c^{-}$ with an $1:1$ ordering of the Mn ions.\cite
{woodward97} However, this model is not able to account for the peak at $%
2\theta \simeq 43.5^{{\rm o}}$, as well as for several other peaks at the
high angles. We were therefore led to rule out this possibility. Recently
Lobanov et al. \cite{lobanov00} reported that the ferromagnetic insulated La$%
_{0.85}$Ca$_{0.15}$MnO$_{3}$ compound undergoes a monoclinic distortion
below the ferromagnetic transition (space group $P2_{1}/c$) with
nonequivalent MnO$_{2}$ layers alternating along the $a$ axis. In their
model they considered that the monoclinic angle is between the $b$ and $c$
axis of the high temperature $Pnma$ structure. From the specific pattern of
Mn-O distances they proposed an unconventional orbital ordering. We tested
also this model but since it does not predict splitting of the $(202)$
reflection it can not account for the low temperature diffraction pattern of La$%
_{0.15}$Ca$_{0.85}$MnO$_{3}$.
%---------------------------------------------------------------------------
\begin{figure}[tbp] \centering%
\includegraphics[angle=0,width=7.5cm]{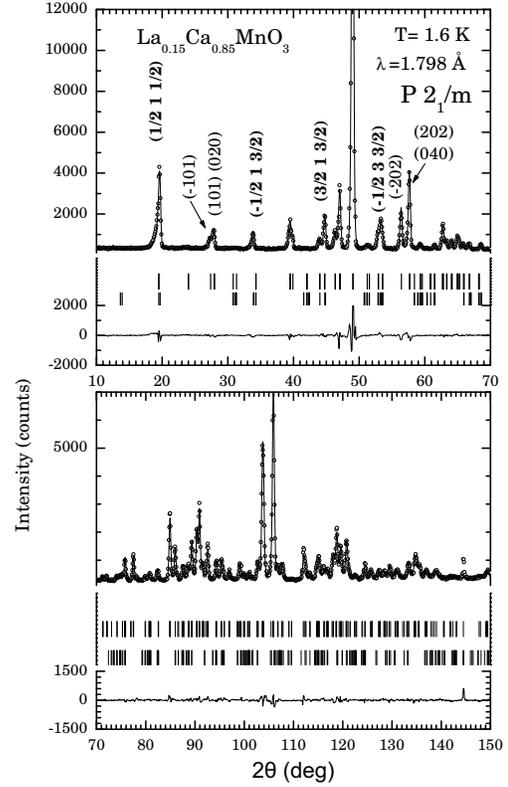}
\caption{
Rietveld refinement pattern for La$_{0.15}$Ca$_{0.85}$MnO$_{3}$ at 1.6 K
($\lambda=1.798$\AA).The observed data points are indicated with open circles,
while the calculated pattern is shown as a continuous line. The
positions of the reflections for the crystal and magnetic structure
employed in the refinements are indicated with vertical lines below the pattern.
}
\label{fig3}%
\end{figure}%
%-------------------------------------------------------------------------
Figure \ref{fig3} shows the 
%Rietveld plot 
observed and calculated neutron powder diffraction intensity patterns
for La$_{0.15}$Ca$_{0.85}$MnO$_{3}$
at 1.6 K (the plot for $x=0.8$ was similar) using neutron
diffraction data from E9 diffractometer. The corresponding structural
parameters for both samples are reported in Table \ref{table1} and the
selected bond distances in Table \ref{table2}. The bonds' length at $T=1.6$
K shows that the Mn(1) site has two long bonds with O2p ($\sim 1.94$ \AA\
for $x=0.85$) in the $a-c$ plane and four short averaging to $\sim 1.88$ \AA.
%Projections of the structure in the $a-c$ and $b-c$ planes at $T=300$ K
%and 1.6 K are illustrated in Fig. \ref{structure}.

In LaMn$^{3+}$O$_{3}$ (which is a pure Mn$^{+3}$ compound) the Mn$^{+3}$O$%
_{6}$ octahedra display a large JT distortion and coherent orbital ordering,
with two long in-plane Mn$^{+3}$O bonds ($2.181$\AA ) and four short ($%
2\times 1.914$ \AA\ in-plane and $2\times 1.966$ \AA\ along the $b$ axis).
\cite{moussa96} Upon substitution of Ca for La in the metallic ferromagnetic
or insulating paramagnetic state the MnO$_{6}$ octahedra have six almost
equal bond length $\sim 1.951$\AA . \cite{dai96} In the case of La$_{0.5}$Ca$%
_{0.5}$MnO$_{3}$ and in the charge-ordered state the structure has two sites
with characteristics of Mn$^{+3}$ and Mn$^{+4}$, respectively. The Mn$^{+3}$O%
$_{6}$ octahedra have two in-plane long bonds $\sim 2.07$ \AA\ and four
short $\sim 1.92$ \AA . On the other hand the Mn$^{+4}$O$_{6}$ octahedra
have six short almost equidistant bonds $\sim 1.91$\AA . \cite{radaelli97}
The same occurs in the La$_{0.33}$Ca$_{0.67}$MnO$_{3}$ compound in the
charge-ordered state, with the Mn$^{+3}$O$_{6}$ octahedra being JT
distorterd ($2\times 2.02$\AA\ and $4\times 1.91$\AA ) and the Mn$^{+4}$O$%
_{6}$ octahedra having five short bonds ($\sim 1.9$\AA ) except for a single
long bond ($\sim 2.0$\AA ) attributed to the frustrated Mn-O-Mn interaction.
\cite{radaelli99}

The main effect of the transition at $T_{c}$ is a redistribution of the Mn-O
bond lengths, rather than that of the bond angles. At room temperature, the
octahedral coordination of manganese with oxygen is approximately
undistorted, with four short bonds $\sim 1.91$\AA\ (two in the basal plane
and two along $b-$axis ) and two longer bonds $\sim 1.92$\AA\ in the basal
plane. Upon cooling below $T_{c}$, the structural changes are evident in the
MnO$_{6}$ octahedra. The two crystallographically independent MnO$_{6}$
octahedra are tetragonally elongated with the two Mn-O$2p$ bonds in the $a-c$
plane elongated and four short bonds (Mn-O$p$ and Mn-O$a)$ along the $b-$%
axis and $a-c$ plane respectively. This is also the manifestation of the
static Jahn-Teller effect. The two manganese sites present in the monoclinic
phase have very similar environments, except for the distortion parameter
(see Table \ref{table2}) which is different by one order of magnitude. 
This fact may imply that the monoclinic phase displays orbital ordering but it does
not show the typical signature of charge ordering. This orbital ordering is
developed concomitantly with the antiferromagnetic ordering (see next
section). Due to absence of large local distortions in the MnO$_{6}$
octahedra one can say that the high temperature phase undergoes a
homogeneous distortion at $T=T_{c}$ and in the low temperature monoclinic
structure the $e_{g}$ electrons are not definitely localized (absence of
charge ordering).
%change
At this point we must note that the charge ordering model can not rull out completely because, 
the difference in the structural parameters for the charge ordering model at the studied consentrations
($x=0.8,0.85$ ) may by small.

The presence of two long bonds for both Mn sites at the monoclinic phase is
a result of the fact that the $e_{g}$ eletrons are $\sigma $-antibonding and
any localization or confinement of them to a specific plane or direction
will result in an expansion of the bonds in that directions. Delocalization
of the $e_{g}$ electrons should lead to a very isotropic distribution of
Mn-O distances in agreement with our observation above $T_{c}$.
%------------------------------------------------
\begin{figure}[tbp] \centering%
\includegraphics[angle=0,width=7.5cm]{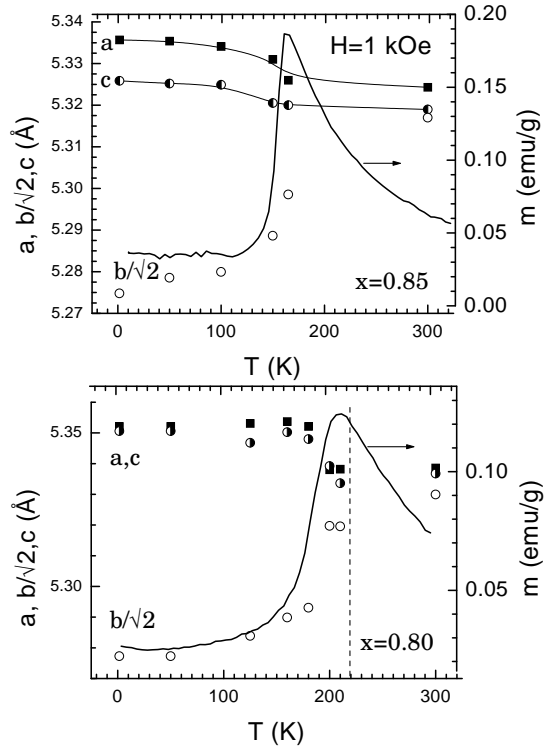}
\caption{
Lattice parameters as a function of temperature for La$_{1-x}$Ca$_{x}$MnO$_{3}$
($x=$0.8 (lower panel) and $x=0.85$ (upper panel)) as obtained from the neutron diffraction data.
Error bars are smaller than the symbols. The bold continuous line represents the magnetic moment
measured after zero field cooling during warming under a $1$ kOe magnetic field.
}
\label{fig4}%
\end{figure}%
%-------------------------------------------------------------------

\begin{figure}[tbp] \centering%
\includegraphics[angle=0,width=7.5cm]{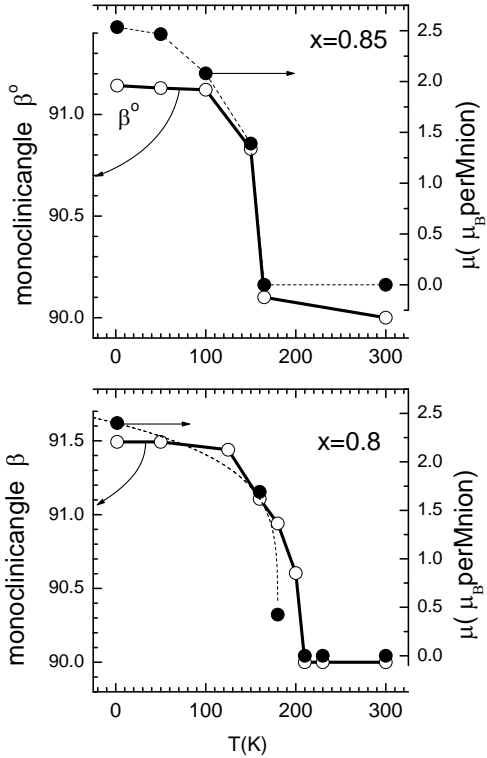}
\caption{
Monoclinic angle $\beta$ and ordered magnetic moment per Mn ion as a function of temperature
for La$_{1-x}$Ca$_{x}$MnO$_{3}$ ($x=0.8, 0.85$) as obtained from the neutron
diffraction data. The continuous lines are guides for the eye.
Error bars are smaller than the symbols.
}
\label{fig5}%
\end{figure}

Figure \ref{fig4} shows the variation of the lattice parameters $a,b/\sqrt{2}
$ and $c$ with temperature for both samples. The lattice parameters display
changes below $T_{\max }$. The $b-$axis undergoes a drastic decreasing while
the $a$ and $c$ axes slightly increase. Since $a>c>b/\sqrt{2}$ the $x=0.8$
and 0.85 samples belong to the so-called $O^{/}$ structure. This type of
structure can result from octahedral tilting and a cooperative JT distortion
(orbital ordering). For comparison we include in this figure the temperature
variation of the magnetic moment deduced from the SQUID measurements during
warming, for both samples (see right scale). Both measurements display a
maximum at $T_{\max }=210$ K and $160$ K for $x=0.8$ and $0.85$ samples,
respectively. The observed magnetization peak can be explained by
considering that at high temperatures the hopping of the $e_{g}$ electron
induces ferromagnetic correlations through the double-exchange mechanism and
when these electrons freeze the ferromagnetic fluctuations are replaced by
superexchange driven antiferromagnetic spin fluctuations.
%-------------------------------------------------
\begin{table}[tbp] \centering%
\caption{
Selected bond lengths (\AA)  and bond angles (deg ) for La$_{1-x}$Ca$_x$MnO$_3$ ($x=0.8,0.85$)
samples at $T=1.6$ and 300 K, calculated according to the structural parameters presented in Table I.
At 300 K (1.6 K) the structural model predict one (two) Mn sites. The bond with subscript '$a$' denotes
bond mainly along the $b$-axis, while the '$p$' stands for bonds in the $a-c$ plane. The table also
includes the average bond length $<d>=(1/6)\sum_{i=1,6}d_i$, the distortion parameter of the
MnO$_6$ octahedra $s^2_d=(1/6)\sum_{i=1,6}[(d_i-<d>)/<d>]^2$.
% and the valence sum $V_i$ for the studied
%compounds, and for LaMn$^{+3}$O$_3$ and CaMn$^{+4}$O$_3$.
}
\begin{ruledtabular}
\begin{tabular}{ccccc}
$x$ & \multicolumn{2}{c}{0.8} & \multicolumn{2}{c}{0.85} \\
$T$ (K) & 300 & 1.6 & 300 & 1.6 \\
\tableline 
Mn-O$a\times 2$ & 1.915(1) & - & 1.908(1) & - \\
Mn-O$p\times 2$ & 1.925(1) & - & 1.923(3) & - \\
Mn-O$p\times 2$ & 1.912(4) & - & 1.908(3) & - \\
Mn(1)-O$1a\times 2$ & - & 1.895(1) & - & 1.894(1) \\
Mn(1)-O$1p\times 2$ & - & 1.886(5) & - & 1.887(1) \\
Mn(1)-O$2p\times 2$ & - & 1.952(7) & - & 1.941(5) \\
Mn(2)-O$2a\times 2$ & - & 1.895(1) & - & 1.898(1) \\
Mn(2)-O$1p\times 2$ & - & 1.920(6) & - & 1.909(4) \\
Mn(2)-O$2p\times 2$ & - & 1.931(6) & - & 1.925(5) \\
Mn(1)-O$1a$-Mn(1) & 159.3(3) & 159.9(5) & 160.0(4) & 159.8(4) \\
Mn(2)-O$2a$-Mn(2) & - & 159.7(5) & - & 158.3(4) \\
Mn(2)-O$1p$-Mn(1) & 159.1(2) & 157.5(4) & 158.3(2) & 158.7(3) \\
Mn(2)-O$1p$-Mn(1) & - & 161.5(3) & - & 159.8(5) \\
\tableline
La-O$p\times 2$ & 2.369(6) & - & 2.344(6) & - \\
La-O$a\times 1$ & 2.37(1)  & - & 2.38 (1) & - \\
La-O$p\times 1$ & 2.527(6) & - & 2.513(6) & - \\
La-O$p\times 2$ & 2.610(6) & - & 2.583(6) & - \\
La-O$p\times 2$ & 2.640(5) & - & 2.651(6) & - \\
La-O$p\times 1$ & 2.864(6) & - & 2.864(6) & - \\
La-O$a\times 1$ & 2.97(1)  & - & 2.94 (1) & - \\
La-O$p\times 2$ & 3.098(6) & - & 3.112(6) & - \\
La(1)-O$2p\times 2$ &-& 2.374(9) &-&  2.36 (1)  \\
La(1)-O$2a\times 1$ &-& 2.38(1)  &-&  2.35 (1)  \\
La(1)-O$1a\times 1$ &-& 2.55 (1) &-&  2.50 (1)  \\
La(1)-O$1p\times 2$ &-& 2.58 (1) &-&  2.583(6)  \\
La(1)-O$2p\times 2$ &-& 2.613(8) &-&  2.595(6)  \\
La(1)-O$1a\times 1$ &-& 2.84 (1) &-&  2.87 (1)  \\
La(1)-O$2a\times 1$ &-& 2.97(1)  &-&  2.98 (1)  \\
La(1)-O$1p\times 2$ &-& 3.122(9) &-&  3.114(7)  \\
La(2)-O$1p\times 2$ &-& 2.34 (1) & - & 2.370(9)  \\
La(2)-O$1a\times 1$ &-& 2.39(1)  & - & 2.38 (1)  \\
La(2)-O$2a\times 1$ &-& 2.53 (1) & - & 2.48 (6)  \\
La(2)-O$2p\times 2$ &-& 2.57 (1) & - & 2.564(7)  \\
La(2)-O$1p\times 2$ &-& 2.705(6) & - & 2.654(7)  \\
La(2)-O$2a\times 1$ &-& 2.87 (1) & - & 2.91 (1)  \\
La(2)-O$1a\times 1$ &-& 2.96(1)  & - & 2.95 (1)  \\
La(2)-O$2p\times 2$ &-& 3.038(9) & - & 3.077(7)  \\
\tableline
$<d_{1}>$ & 1.917(2) & 1.911(2) & 1.913(2) & 1.907(2) \\
$<d_{2}>$ &  & 1.915(2) &  & 1.910(2) \\
$<s_{d_{1}}^{2}\times 10^{-5}$ & 1.61 & 44.68 & 2.6 & 30.14 \\
$<s_{d_{2}}^{2}\times 10^{-5}$ &  & 7.95 &  & 6.45 \\
%$V_{1}$ & 3.84 & 3.92 & 3.89 & 3.96 \\
%$V_{2}$ &  & 3.87 &  & 3.91 \\
%$V($Mn$^{+3})$ & 3.10 &  &  &  \\
%$V($Mn$^{+4})$ & 4.04 &  &  &
\end{tabular}
\end{ruledtabular}
\label{table2}%
\end{table}%
%--------------------------------------------------------------------------
Figure \ref{fig5} shows the variation of the monoclinic angle $\beta $ as a
function of temperature for both samples. In the same plot we include the
temperature variation of the ordered magnetic moment per Mn ion (deduced
from the magnetic peak intensity in neutron diffraction, see next section)
for both samples. The monoclinic angle increases from 90$^{{\rm o}}$ at $%
T_{c}$ to $\sim 91^{{\rm o}}$ at low temperatures and follows the
temperature variation of the ordered magnetic moment shown in fig. \ref{fig5}%
. Also, the transition point marked by the appearance of the monoclinic
distortion coincides with the peak observed in the SQUID data (fig. \ref
{fig4}). The space group describing the low temperature structure is a
maximal subgroup of the high temperature phase. Then, according the Landau
theory\cite{landau} of phase transitions one expect that the transition is
of second order. In other words it is possible to go from the $Pnma$
structure to $P2_{1}/m$ by a continuous change in the cation displacements.

Finally, we discuss the possibility that the Mn(2) sites are occupied
exclusively by Mn$^{+4}$ ions. The Mn(1) sites with longest bond length $%
\sim 1.95$\AA\ ($x=0.80$ sample) corresponds to a site with a Mn$^{+3}$
character. For example it can be considered as randomly occupied by $60\%$ Mn%
$^{+4}$ and $40\%$ Mn$^{+3}$. Of course the Mn$^{4+}$-O bond length is
shorter by $\sim 0.05$\AA\ in comparison with $x=0.5$ or $2/3$. The
distortion is smaller for the sample with $x=0.85$ than for that with $x=0.80
$. Therefore, we might conclude that if at $x=0.75$ a clear charge-ordered
state exists then when Ca content is further increased, the crystal is
tending more and more to acquire the structure of CaMnO$_{3}$ with a
homogeneous distortion of the $T=300$ K $Pnma$ phase.

\subsection{Magnetic structure}

In this section we will discuss the magnetic reflections appearing at $%
T<T_{c}$ using neutron diffraction data from the E6 diffractometer. Figure
\ref{fig2} shows the neutron diffraction patterns of the La$_{0.15}$Ca$%
_{0.85}$MnO$_{3}$ sample recorded with $\lambda =2.44$\AA\ in the range $%
1.6-165$ K. Below $T_{c}=150$ K splitting of the family of peaks [(101),
(020)] and (111) due to the monoclinic distortion is observed. A series of
additional Bragg peaks occurring mainly at low angles give a clear
indication for the presence of antiferromagnetic long-range ordering. The
antiferromagnetic ordering occurs concomitantly with the monoclinic
transition at $T_{c}$ ($T_{N}=T_{c}$). All the magnetic peaks can be indexed
using the propagation vector ${\bf k}=[\frac{1}{2},0,\frac{1}{2}]$ ($U$
point of the first Brillouin zone). The magnetic structure for this
propagation vector corresponds to the so-called type-C antiferromagnetic
structure. \cite{wollan55,radaelli99}
%------------------------------------------------
%-------------------------------------------------

The integrated intensity of a magnetic reflection at ${\bf q=Q+k}$ (${\bf Q}$
is a reciprocal lattice vector) for a colinear magnetic structure in a
neutron powder diffraction pattern can be written as:
\begin{equation}
I({\bf q})=\frac{I_{0}}{V^{2}}\frac{1}{\sin \theta \sin 2\theta }%
m\left\langle 1-(\widehat{{\bf q}}\widehat{\cdot {\bf s}})^{2}\right\rangle
\left| F({\bf q})\right| ^{2}  \label{eq1}
\end{equation}
where $I_{0}$ is the scale factor, $V$ is the unit cell volume, $\theta $ is
the Bragg angle and $m$ is the multiplicity of the reflection ${\bf q}$. The
term $\left\langle 1-(\widehat{{\bf q}}\widehat{\cdot {\bf s}}%
)^{2}\right\rangle $ is an average over all the equivalent $\{{\bf q}\}$
reflections, $\widehat{{\bf q}}$ is the unit scattering vector, $\widehat{%
{\bf s}}$ is the unit vector along the axis of the colinear magnetic
structure and $F({\bf q})$ is the magnetic structure factor for the
configurational symmetry. The magnetic structure factor for the particular
crystal and magnetic structure can be written as
\begin{widetext}
\begin{equation}
F({\bf q}) =\left\{ 1-\exp (\pi ik)\right\}  
\left\{ p_{1}\exp \left[ \pi i\left( h+\frac{1}{2}\right) \right]
+p_{2}\exp \left[ \pi i\left( l+\frac{1}{2}\right) \right] \right\}
\label{eq2}
\end{equation}
\end{widetext}
where $p_{j}=(0.269\times 10^{-12}{\rm cm/\mu }_{B})\times S_{j}\times
f_{j}\times \exp (-W_{j})$ , $S_{1}$ ($S_{2}$) is the average ordered
magnetic moment in Bohr magnetons $\mu _{B}$ for the Mn ion at (1/2,0,0)
site \ ((0,0,1/2)), $f_{j}$ is the magnetic form factor for the Mn ion and $%
W_{j}$ is the Debye-Waller factor for the $j$th Mn ion. Thanks to the
monoclinic distortion the diffraction pattern revealed zero intensity for
the resolved reflections $\left( \frac{1}{2},\pm 1,-\frac{1}{2}\right) $ and
$\left( \frac{3}{2},\pm 1,\frac{1}{2}\right) ,\left( \frac{1}{2},\pm 1,\frac{%
3}{2}\right) .$ Using Eq. \ref{eq2} we can readily conclude that $p_{1}=p_{2}
$. The ferromagnetic interactions occur along the two long in-plane Mn-O
bonds, while along the four short Mn-O bonds the interactions are
antiferromagnetic. The relative intensity of the reflections $\left( \frac{1%
}{2},\pm 1,\frac{1}{2}\right) $ and $\left( -\frac{3}{2},\pm 1,\frac{1}{2}%
\right) ,\left( -\frac{1}{2},\pm 1,\frac{3}{2}\right) $ implies that ${\bf %
S\perp k}$, that is, the magnetic moment is directed along the two long
in-plane Mn-O bonds.The fit of the experimental pattern at $T=1.6$ K gives
an amplitude for the ordered magnetic moment per Mn ion of $2.53(2)$ and $%
2.47(1)$ $\mu _{B}$ for $x=0.8$ and $0.85$, respectively. These values are
smaller than the theoretically expected for the stoichiometric mixtures of Mn%
$^{+4}$ and Mn$^{+3}$ (e.g. $0.15\times 4+0.85\times 3=3.15)$. The smaller
observed magnetic moment than their nominal value of random mixture indicate
the hybridization of the Mn $t_{2g}$ orbitals and the O $2p$ orbitals. The
temperature variation of the ordered magnetic moment per Mn ion as obtained
from Rietveld refinement of the neutron powder diffraction data for both
samples is plotted in Fig \ref{fig4}.

\section{Conclusions}

The present work based on the analysis of neutron powder diffraction data
shows that $x=0.80$ and $x=0.85$ undergo an orthorombic-to-monoclinic
structural transition around $210$ K and $160$ K, respectively. The
monoclinic distortion is larger for the $x=0.80$ sample. Simultaneously,
long range antiferromagnetic order is developed and it can be described with
propagation vector ${\bf k}=[\frac{1}{2},0,\frac{1}{2}]$ that corresponds to
the C-type structure. The ordered magnetic moment of the Mn ions is found to
be normal to the propagation vector.

A key question was about the existence or not of charge-ordering at low temperatures. 
Charge-ordering models without or with monoclinic distortions have failed to refine 
the diffraction patterns. The refinement was successful only with the $P2_1/m$ space
group, where the two Mn sites in the low temperature structure have similar environments
(in terms of their bond lengths in the corresponding MnO6 octahedra). 
This absence of preferential local distortions (which would be a sign of different charges
in the two Mn sites) in the MnO6 octahedra have led us to conclude that charge-ordering
does not occur in the monoclinic phase. However, the difference in the structural parameters
for charge-order and charge disorder of Mn$^{+3}$ and Mn$^{+4}$ with ratios of $0.15:0.85$
and $0.2:0.8$ may be as small (or smaller) as the sensitivity of the experimental method.

%The two Mn sites in the low temperature structure have similar environments
%(in terms of their bond lengths in the corresponding MnO$_{6}$ octahedra).
%Based therefore on the absence of preferential local distortions of the MnO$%
%_{6}$ octahedra, reflected also in the results of bond valence sum
%calculations, it seems reasonable to conclude that charge-ordering does not
%occur in the monoclinic phase.

\acknowledgements{
This work was partial supported from the Greek Secretariat for
Research and Technology through the PENED program 99ED186 and
by the EC through the CHRX-CT93-0116 access to large-scale fa-
cilities project.
}

\end{document}